\documentclass[12pt]{article}
\usepackage{graphicx}
\usepackage{epsfig}
\usepackage{epstopdf}
\usepackage{amsmath}
\usepackage{amsfonts}
\usepackage{amssymb}
\usepackage{ccaption}
\usepackage[usenames]{color}

\usepackage[
      colorlinks=true,
      linkcolor=blue,  
      citecolor=blue,
      urlcolor=blue,    
      filecolor=blue,     
      linkcolor=blue,
      pdfstartview=FitV,
      pdftitle={},
        pdfauthor={Hubeny},
        pdfsubject={bulk probes},
        pdfkeywords={AdS/CFT},
        pdfpagemode=None,
        bookmarksopen=true    
      ]{hyperref}

\definecolor{labelkey}{gray}{0.1}
  \captionnamefont{\bfseries}
  \captiontitlefont{\small\sffamily}
  \captiondelim{: }
  \hangcaption

\def\sec#1{Section \ref{#1}}
\def\fig#1{Fig.\,\ref{#1}}
\def\req#1{(\ref{#1})}
\def\App#1{Appendix \ref{#1}}

\def\ands{\qquad {\rm and} \qquad}

\def\({\left(}
\def\){\right)}

\def\CE{{\cal E}}

\def\CL{{\cal L}}

\def\CN{{\cal N}}
\def\CO{{\cal O}}

\def\transv{{\vec V}_{\perp}}

\def\GSW{\{GSW\}}

\def\vev#1{\langle\, #1 \, \rangle}
\def\expval#1{{\langle \, #1  \, \rangle}}

\numberwithin{equation}{section}

\begin{document}

\setlength{\unitlength}{1mm}

\begin{titlepage}

\begin{flushright}
DCPT-10/75
\end{flushright}
\vspace{1cm}

\begin{center}
{\bf \Large Holographic dual of collimated radiation}
\end{center}

\vspace{4mm}
\begin{center}
Veronika E.~Hubeny$^{a}$

\vspace{.4cm}
{\small \textit{$^{a}$Centre for Particle Theory \& Department of Mathematical Sciences, Durham University, }}\\
{\small \textit{Science Laboratories, South Road, Durham DH1 3LE, United Kingdom}} \\
\vspace*{0.4cm}
{\small {\tt veronika.hubeny@durham.ac.uk}}
\end{center}

\vspace{5mm}

\begin{abstract}

We propose a new and simple method of estimating the radiation due to an accelerated quark in a strongly coupled medium, within the framework of the AdS/CFT correspondence. 
In particular, we offer a heuristic explanation of the collimated nature of synchrotron radiation produced by a circling quark,
which was recently studied in
 Phys.Rev.{\bf D81} (2010) 126001.
The gravitational dual of such quark is a coiling string in AdS, whose backreaction on the spacetime geometry remains tightly confined, as if `beamed' towards the boundary.
While this appears to contradict
conventional expectations from the scale/radius duality,
we resolve the issue by observing that 
 the backreaction of a relativistic string is reproduced by a superposition of gravitational shock waves.  We further demonstrate that this proposal allows us to reduce the problem of computing the boundary stress tensor to merely calculating geodesics in AdS, as opposed to solving linearized Einstein's equations.

\end{abstract}

\end{titlepage}

\tableofcontents
\newpage

\section{Introduction}
\label{s:intro}

Since its inception 13 years ago, the gauge/gravity duality \cite{Maldacena:1997re,Gubser:1998bc,Witten:1998qj} has engendered vigorous exploration of properties of non-abelian gauge theories by studying their gravitational dual.  One of the crucial attributes is the strong/weak coupling nature of the duality: in the limit of strong coupling and high rank of the gauge group, the theory is described in terms of classical gravity on a higher-dimensional asymptotically Anti de Sitter (AdS) background.
Such dual picture has offered invaluable insight into the behavior of the field theory in the regime of strong coupling where quantum effects dominate and traditional methods prove inadequate.

One of the most natural questions concerning properties of a strongly-coupled medium concerns its ability to transmit signals.  In particular, how does `radiation' propagate in such a medium?  While this question has been well-studied for numerous weakly coupled systems, the strong coupling case is much more interesting, and of potential relevance to experimentally attainable settings.  Of course, to achieve closest contact with systems such as the quark gluon plasma produced at the Relativistic Heavy Ion Collider, one would like to study strongly coupled field theories at high temperature.  

Nevertheless, while perhaps less indicative of real-world systems, the zero-temperature setting is also of active interest, since it provides an excellent toy model for the more realistic cases while being amenable to exact calculations.
From the gravitational dual standpoint, zero temperature constitutes a drastic simplification because the vacuum of the gauge theory corresponds to pure AdS, whereas a thermal state corresponds to a Schwarzschild-AdS black hole, which is both geometrically and causally more complicated.  For example, while an exact solution for a string with its endpoint following a given trajectory is known in pure AdS for arbitrary endpoint trajectories, in Schwarzschild-AdS an analogous solution is known only in special cases with high enough symmetry.  Nevertheless, once a given setup has been well-understood at zero temperature, one can then use this as a crutch to extend the analysis to non-zero temperature. 

A very interesting recent study of radiation in strongly coupled gauge theory at zero temperature was undertaken in \cite{Athanasiou:2010pv} for the case of a quark in uniform circular motion.\footnote{
In particular, \cite{Athanasiou:2010pv} consider a test quark (modeled by an infinitely massive spin $\frac{1}{2}$ particle of an $\CN = 2$ hypermultiplet), coupled to high gauge group rank ($N_c \to \infty$) limit of 3+1 dimensional Super Yang-Mills, at finite but large 't Hooft coupling $\lambda \gg 1$.  
}
The gravitational dual consists of a fundamental string in AdS spacetime whose endpoint is anchored at the quark on the boundary.  The quark's trajectory then determines the string's embedding in AdS.\footnote{
In \cite{Mikhailov:2003er} the string's embedding was obtained explicitly in terms of (arbitrary) quark trajectory in a certain gauge (using worldsheet retarded time), whereas \cite{Athanasiou:2010pv} give the answer for the circular motion directly in Poincare coordinates.
}
The string, having non-trivial stress tensor supported on its world-volume, backreacts on the bulk spacetime, which in turn induces a non-trivial stress tensor in the boundary field theory.
The authors of \cite{Athanasiou:2010pv} compute the corresponding boundary energy density by solving linearized Einstein's equations in the bulk.  
They find a rather surprising result: the radiation closely resembles synchrotron radiation exhibited in classical electrodynamics and weakly-coupled SYM.  In particular, the radiation pulse does not broaden as it propagates outward, despite the field theory being strongly coupled.\footnote{
Recently, \cite{Hatta:2010dz}
argued this to be a generic result, pertaining to arbitrary trajectory at zero temperature. Their viewpoint is however very different from ours, as they ascribed the absence of broadening to the string's backreaction being sourced only near the boundary.}

Indeed, this result is just as remarkable from the gravitational dual standpoint, since it implies that the backreaction effects from string bits deep in the bulk are somehow `beamed' towards the boundary, contradicting naive expectations from the scale/radius (or UV/IR) duality \cite{Susskind:1998dq}, according to which an excitation deeper in the bulk should be represented by a more dispersed signal in the boundary field theory.  This mysterious beaming effect in fact prompted the present study, which was recently summarized in \cite{Hubeny:2010rd}.  
In particular, our main goal is to attain better geometrical understanding of how the backreaction effects of a bulk string manage to produce a tightly collimated beam on the boundary.  As it turns out, this has the added benefit of simplifying the stress tensor calculation itself.

While far simpler than any conceivable direct method within the field theory, the calculation of \cite{Athanasiou:2010pv} is still fairly involved.  Here we propose a much easier method to obtain the boundary stress tensor, in a straightforward and computationally non-intensive manner which essentially reduces to finding geodesics in AdS.  This simplification applies to the regime where the string is sufficiently relativistic, which, as we discuss below, includes the present case of interest, and indeed occurs quite generically for bulk dual describing radiation from an accelerating quark.  It rests on the observation that in such a regime, the backreaction of the string is captured by a superposition of gravitational shock waves sourced by individual string bits.   
Justifying and testing this proposal against the results of \cite{Athanasiou:2010pv} will be the main focus of the present work.

Once the string's backreaction is understood in terms of gravitational shock waves, it elucidates the mechanism producing the observed non-dispersive character of the boundary radiation.  This is due to the geometrical nature of the gravitational shock waves \cite{Aichelburg:1970dh,Dray:1984ha}: each shock wave is supported on a null hypersurface in the spacetime, which means that the influence of the backreaction is not allowed to disperse as naively expected.  As we will demonstrate, it is the combined effect from the set of all such shock waves produced by the string that leads to the observed phenomenon of the radiation pulses having correct width and propagating at the speed of light without broadening.  This has further interesting implications for the scale/radius duality, as discussed below.

The plan of this paper is as follows.  In \sec{s:setup} we introduce the circling quark configuration and the corresponding bulk dual in more detail.  This allows us to review the key observations of \cite{Athanasiou:2010pv}, which provide a sharp testing-ground for our proposal.  In \sec{s:beaming} we detail our proposal for the beaming mechanism.  We first justify our expectations based on general principles and then explain how to calculate the backreaction explicitly.  The actual calculations are relegated to the Appendices, in order to maintain the flow of the main text.   Having outlined the construction, in \sec{s:results} we present the results of implementing this construction.  We first do so numerically, by plotting various aspects of our prediction for the energy density induced on the boundary.  We then determine the salient features of this energy density analytically in the asymptotic regime, which allows more detailed quantitative comparison with \cite{Athanasiou:2010pv}.  We summarize our findings in \sec{s:extensions} and end by discussing further applications and extensions of our construction.

\section{Set-up: bulk dual for circling quark}
\label{s:setup}

Let us start by reviewing the essential ingredients of the work of 
 \cite{Athanasiou:2010pv}, which uses the AdS/CFT correspondence to study the radiation from a quark in uniform circular motion in a strongly coupled field theory at zero temperature.
 In the gravitational dual, such set-up corresponds to a string in AdS, 
 ending on the quark.  
As indicated above, the gravitational backreaction of the string deforms the bulk geometry away from AdS, and this in turn induces a non-trivial stress tensor on the boundary, which captures both the motion of the quark, as well as the resulting radiation.  

To analyze the set-up more explicitly, let us first fix the notation.
We will consider the field theory to be formulated on Minkowski spacetime\footnote{
As a matter of convenience, in the following we will use both Cartesian and spherical coordinates with the boundary metric 
$ds_{\rm bdy}^2 = - dt^2 + dx^2 +dy^2 +dz^2 =
- dt^2 + dr^2 + r^2  \left( d\theta^2 + \sin^2 \theta \, d \varphi^2 \right) $.
}  
 so the notion of radiation is best-defined in the asymptotic region $r \to \infty$.
The source of the radiation is a test quark moving on a circle of radius $R_0$ with constant angular velocity $\omega_0$, so that 
its trajectory is given by
\begin{equation}
r(t)=R_0,\quad \theta(t)=\frac{\pi}{2},\quad \varphi(t)=\omega_0 \, t \ ,
\label{}
\end{equation}
and the quark moves with constant speed $v=R_0 \, \omega_0$, with the corresponding Lorentz factor $\gamma \equiv \frac{1}{\sqrt{1-v^2}}$.
In the gravitational dual, this set-up corresponds to a trailing string moving in AdS.  In the Poincare (or Fefferman-Graham) coordinates  $(t,r,\theta,\varphi,u)$, where  AdS (of unit size) has the metric
\begin{equation}
ds^2 = \frac{1}{u^2} 
\left[ - dt^2 + dr^2 + r^2  \left( d\theta^2 + \sin^2 \theta \, d \varphi^2 \right) + du^2 \right] \ ,
\label{AdSmet}
\end{equation}	
$u$ corresponds to the bulk radial coordinate, with $u=0$ representing the AdS boundary.
Parameterizing the string worldsheet by $t$ and  $u$, \cite{Athanasiou:2010pv} write the string embedding as 
\begin{equation}
X^{M}(t,u) = (t,R(u), \frac{\pi}{2},\phi(u) +\omega_0 \, t , u) \ , 
\qquad
\label{Xcoords}
\end{equation}	
and solve the Nambu-Goto action $S_{NG} = -T_0 \, \int \, dt \, du \, \sqrt{-\gamma} $, which for the induced metric on the worldsheet
$\gamma_{ab} = \partial_a X^M \, \partial_b X^N \, g_{MN}$ with $g_{MN}$ given by \req{AdSmet} becomes
\begin{equation}
S_{NG} = - \frac{\sqrt{\lambda}}{2 \pi} \, \int \, dt \, du \, 
\frac{\sqrt{(1-\omega_0^2 \, R^2) \, (1+R'^2)+R^2 \, \phi'^2}}{u^2} \ ,
\label{}
\end{equation}	
to determine the profile functions $R(u)$ and $\phi(u)$: 
\begin{equation}
R(u) = \sqrt{R_0^2 + v^2 \, \gamma^2 \,  u^2}  \qquad {\rm and} \qquad
\phi(u) = - v  \, \gamma  \, \frac{u}{R_0} + \tan^{-1} \left(  v  \, \gamma  \, \frac{u}{R_0}  \right) \ .
\label{string}
\end{equation}	
From these we can see that the string has the shape of a  rigidly rotating helix, flaring-out into the bulk at a rate $R'(u) \approx v  \, \gamma$.
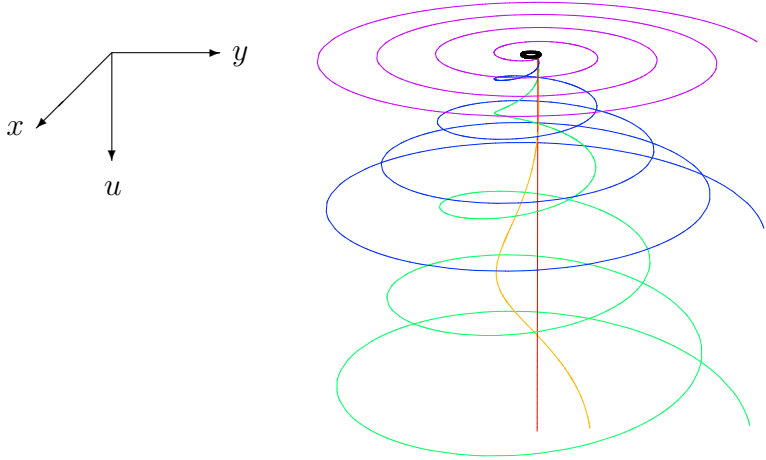
\begin{figure}
\begin{center}
\includegraphics[width=2.5in]{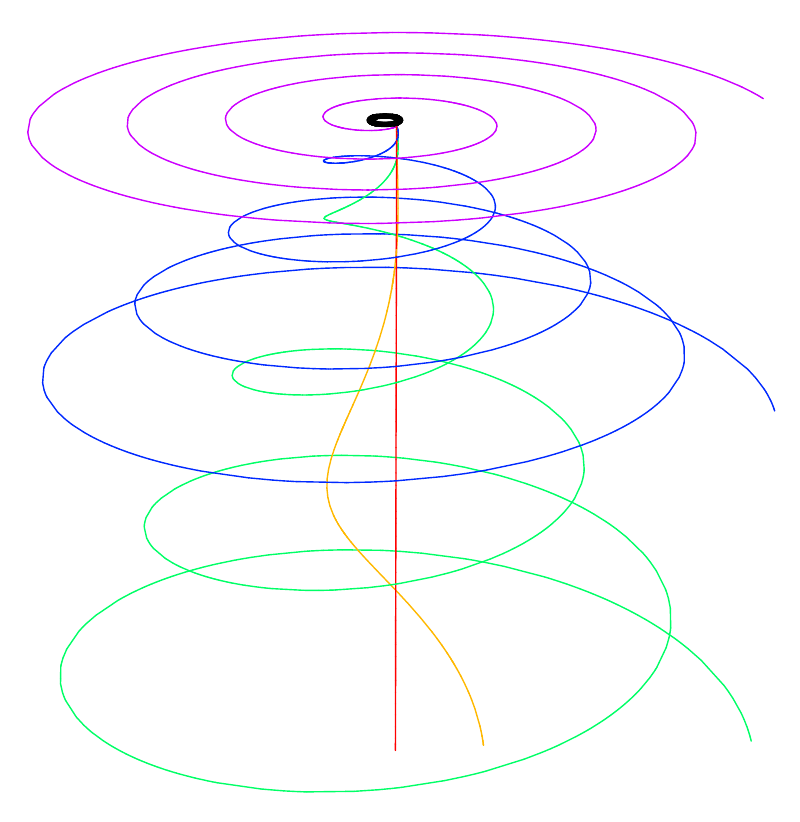}
\begin{picture}(0,0)
\setlength{\unitlength}{1cm}
\put(-9,5.6){{\vector(1,0){1.44}}}
\put(-9,5.6){{\vector(-1,-1){1}}}
\put(-9,5.6){{\vector(0,-1){1.44}}}
\put(-9.1, 3.7){$u$}
\put(-10.4, 4.5){$x$}
\put(-7.4, 5.5){$y$}
\end{picture}
\caption{The spiral string solution in AdS$_5$ spacetime corresponding to quark undergoing a circular motion. Several curves are drawn, corresponding to different quark velocities $v$; specifically, $v=0$ (vertical red line), $0.15$ (yellow), $0.5$ (green), $0.8$ (blue), and $0.99999$ (horizontal purple spiral).  The small black circle on the top indicates the quark's trajectory on the AdS boundary; the vertical direction corresponds to the bulk radial direction.}
\label{f:bulkstring}
\end{center}
\end{figure}
This is illustrated in \fig{f:bulkstring} for several values of the quark velocity $v$.  We see that in the limit $v \to 1$, the string hugs the boundary, so that we expect the boundary stress tensor being localized just above the string, as is indeed confirmed by explicit calculation. 
We now discuss several key aspects of the geometry in a bit more detail. 

\paragraph{Worldsheet horizon:}
Let us first consider the causal structure of the worldsheet metric.
It is intriguing to note that despite the bulk having the causally trivial structure of pure AdS, the  worldsheet metric corresponds to  a black hole geometry,\footnote{
This is by now a familiar result (see e.g.\ \cite{Chernicoff:2008sa, Xiao:2008nr,  Chernicoff:2009re, Chernicoff:2010yv}).
} with an event horizon at
\begin{equation}
u_h = \frac{R_0}{v^2 \, \gamma^2} \ .
\label{WShor}
\end{equation}	
(This is easy to see by finding the induced metric on the worldsheet
 $\gamma_{ab}$ explicitly, 
\begin{equation}
\gamma_{tt} = - \frac{1-\omega_0^2 \, v^2 \, \gamma^4 \, u^2}{\gamma^2 \, u^2} \ , \qquad
\gamma_{tu} = - \omega_0^2 \, v^2 \, \gamma^3  \ , \qquad
\gamma_{uu} = \frac{1+\omega_0^2 \, v^2 \, \gamma^4 \, u^2}{u^2}  \ ,
\label{}
\end{equation}	
and noting that since the metric is static, the Killing horizon at $\gamma_{tt}=0$ coincides with the event horizon.)
 This suggests that beyond $r \approx R(u_h) = \frac{R_0}{v}$, the boundary stress tensor will describe radiation form the quark rather than the gluonic `halo' of the quark itself.
The total power radiated by the quark coincides with the energy flux down the string, which \cite{Athanasiou:2010pv} evaluate to be $P = \frac{\sqrt{\lambda}}{2 \pi} \, a^2$, where $a = v^2 \, \gamma^2 /R_0  = 1/u_h$ is the quark's proper acceleration.

\paragraph{Transverse velocity:}
An important quantity in what follows is the string's velocity. 
 Since a fundamental string has no longitudinal excitations, a natural quantity to describe the string's motion is the transverse velocity of the string, $\transv(t,u)$.  
In the static gauge chosen above with $X^M = (t, \vec X)$, $\transv$ is a purely spatial vector, given by the component of $\frac{\partial \vec X}{\partial t} $ perpendicular to the string, 
\begin{equation}
\transv(t,u) \equiv \vec \tau - 
\left( \frac{\vec \tau \cdot \vec \sigma}{\vec \sigma \cdot \vec \sigma} \right) \vec \sigma
\ , \qquad {\rm where} \ \
\vec \sigma \equiv \frac{\partial \vec X}{\partial u} \ , \ \ 
\vec \tau \equiv \frac{\partial \vec X}{\partial t} \ .
\label{transvform}
\end{equation}	
The time dependence is trivial since the string rotates rigidly,
and using the parameterization \req{Xcoords}, in the coordinates $(r,\phi,u)$, we find that 
\begin{eqnarray}
\transv(u) &=& \frac{-\omega_0 \, R(u)^2 \, \phi'(u)}{R'(u)^2+R(u)^2 \, \phi'(u)^2 + 1 }
\, \left( R'(u) \ ,\  -\frac{R'(u)^2 + 1}{R(u)^2 \, \phi'(u)} \ ,\  1 \right) \cr 
&&\cr 
&=&  \frac{v^4 \, \gamma^3 \, u^2}{ R_0^2+ v^4 \, \gamma^4 \, u^2 } \, 
\left( \frac{v^2 \, \gamma^2 \, u}{\sqrt{R_0^2+  v^2 \, \gamma^2 \, u^2}}
 \ ,\  
 \frac{R_0}{v^3 \, \gamma^3 \, u^2} \, \frac{ R_0^2+ v^2 \, \gamma^4 \, u^2 }{ R_0^2+ v^2 \, \gamma^2 \, u^2 }
  \ ,\  1
\right) \ .
\label{transv}
\end{eqnarray}	
We see that the direction of the transverse velocity points towards larger $r$, $\phi$, and $u$. 
In particular, note that $\transv$ points {\it away} from the AdS boundary.
In the interesting regime of large $\gamma$ and $u/R_0$, the leading $(r,\phi,u)$ components of $\transv{}$ approach $\left( 1 ,\frac{R_0}{\gamma^2 \, u^2} ,\frac{1}{\gamma} \right)$.

Of greater interest is the magnitude of the transverse velocity.  Using \req{transv} we can evaluate\footnote{
Note that in taking the dot product $V_{\perp}^2 = \transv{} \cdot \transv{}$, we use only the flat spatial metric without its conformal factor $\frac{1}{u^2}$ because the $t$ direction is likewise scaled by the same conformal factor.}
 this to be
\begin{equation}
||\transv(u)|| = \sqrt{\frac{\omega_0^2 \, R(u)^2 \, (R'(u)^2 + 1)}{R'(u)^2+R(u)^2 \, \phi'(u)^2 + 1 }} \ = \ 
  \sqrt{\frac{ v^2 \, R_0^2 + v^4 \, \gamma^4 \, u^2}{R_0^2 + v^4 \, \gamma^4 \, u^2}} \ .
\label{transvel}
\end{equation}	
It is then easy to see that the transverse speed  of the string increases monotonically with $u$, and asymptotes  the speed of light deep in the bulk.
For future use, also note that the product of this speed and the associated Lorentz boost factor $\gamma_{\perp}$, which characterizes to what extend is the string moving relativistically, 
is given by
\begin{equation}
V_{\perp}^2 \, \gamma_{\perp}^2 = \gamma^2 \, (v^2+u^2/u_h^2) \ .
\label{transvgamma}
\end{equation}	
with the worldsheet horizon $u_h$ defined in \req{WShor}.

\paragraph{Boundary stress tensor:}
We now summarize the key results of \cite{Athanasiou:2010pv}.
To find how the radiation induced by the quark's acceleration propagates through the strongly coupled medium, 
\cite{Athanasiou:2010pv} consider the energy density $\CE \equiv T^{tt}$, where $T^{\mu\nu}$ is the boundary stress tensor.  Since the on-shell bulk gravitational action is the generating functional for the boundary stress tensor, $T^{\mu\nu}$ can be read-off from the asymptotic form of the bulk metric deformation.  In particular, using the bulk stress tensor produced by the string, \cite{Athanasiou:2010pv} solve the linearized Einstein's equations near the boundary.
This yields an analytic few-line expression for $\CE(t,r,\theta,\varphi)$  (see eq.(3.70) of \cite{Athanasiou:2010pv}, plotted in their Fig.4) in terms of retarded time $t_{\rm ret}$ which is evaluated numerically. 

As remarked above, this boundary energy density turns out to have rather surprising properties:  it is mainly supported on a thin spiral  rotating rigidly at constant angular velocity, with the energy pulses {\it not} widening in the boundary-radial direction.\footnote{
This statement holds only at $T=0$; at non-zero temperature, these pulses would widen, slow down, and eventually thermalize.  However, for $\omega_0^2 \, \gamma^3 \gg \pi^2 \, T^2$, the synchrotron radiation persists on length scales smaller than the thermal scale $1/T$.} 
The radiation pattern is in fact in close quantitative agreement with that of a corresponding radiation at weak coupling, which likewise mimics that of synchrotron radiation in classical electrodynamics.

More specifically, the energy density $\CE$ exhibits the following salient features.
\begin{itemize}
\item Each pulse remains constant width as it propagates outward, i.e.\ it does not broaden.  
\item A pulse propagates outward at the speed of light, independently of the quark velocity $v$. 
\end{itemize}
If we change the quark velocity $v$, the quatitative features of $\CE$ change in response:
\begin{itemize}
\item The spatial separation between the spiral arms of $\CE$ at a fixed time narrows with increasing $v$ as $L\sim 2 \pi \, R_0 / v$, as implied by the previous point.
\item Pulse width decreases with increasing $v$ as $\sim 1/\gamma^3$.
\item $\CE(\theta)$ decreases off the orbital plane, with the characteristic width narrowing for increasing $v$ as $\sim 1/\gamma$.
\end{itemize}
As argued in \cite{Athanasiou:2010pv}, these features, most notably the first point, are quite surprising from the field theory point of view: since the radiation is strongly coupled, one might have expected it to diffuse and isotropize.  Nevertheless, we will see that in fact all of these features are reproduced by our simple method of estimating the backreaction of a relativistic string, which we now discuss.

\section{Proposal for beaming mechanism}
\label{s:beaming}

The curious features of the radiation propagating through a strongly coupled medium 
summarized above are no less surprising and fascinating from the bulk standpoint: why does the gravitational backreaction of the bulk string behave in such a sharply-localized fashion?  As pointed out in the Introduction, this is at odds with the naive expectations from the scale/radius duality that the deeper parts of the string would produce more diffuse signals on the boundary.  In this section, we propose a general mechanism which leads to the observed localization; we refer to this mechanism as `beaming'. 

First of all, note that large part of the string is moving relativistically.  This is evident from \req{transvel}, and more conveniently quantified by \req{transvgamma}: as long as $v \, \gamma \gtrsim 1$ (for any $u$) or $u \gtrsim u_h$ (for any $v$), the string's transverse velocity is relativistic $V_{\perp}^2 \, \gamma_{\perp}^2  \gtrsim 1$.
The object then is to understand how such a relativistic string backreacts on the spacetime.  We use a well-known fact that the backreaction of a massless particle moving at the speed of light is given by a {\it gravitational shock wave} \cite{Aichelburg:1970dh,Dray:1984ha}.
Such metric perturbation is supported\footnote{
The actual construction of a gravitational shock wave in AdS, discussed in \sec{s:GSWconstruct}, can be obtained similarly to \cite{Aichelburg:1970dh,Dray:1984ha}, by taking a double-scaling limit of a boosted black hole, with the boost parameter taken to infinity and mass to zero, keeping the total energy fixed \cite{Hotta:1992qy,Horowitz:1999gf}.} on a null plane transverse to the particle's velocity, analogously to the behaviour of the electric field for an infinitely-boosted charge in classical electrodynamics.
Accordingly, the backreaction of all `string bits' (viewed as a continuum of particles distributed along the string), moving relativistically in the direction of $\transv$, should be given by a superposition of such gravitational shock waves, one for each particle.
We therefore propose that:
\begin{quote}
{\it The backreaction of the full string is well-approximated by a linear superposition of gravitational shock waves 
normal to the string's transverse velocity.}
\end{quote}
Making this proposal more explicit and testing it will be the main focus of this work.
For compactness of notation, we will henceforth use GSW to abbreviate ``gravitational shock wave" and denote the superposition of all gravitational shock waves sourced by the string as \GSW. 
 In the remainder of this section, we first justify why \GSW\ is a good approximation to the string's backreaction, we then explain how to find the \GSW\ explicitly, and finally we indicate how to extract the boundary stress tensor induced by such \GSW.  Since the exact result for the boundary energy density $\CE$ is known from the calculation of \cite{Athanasiou:2010pv}, we can readily test our proposal by comparing our results with \cite{Athanasiou:2010pv}, which will be carried out in detail in \sec{s:results}.

\subsection{Justification for \GSW}
\label{s:justification}

Our proposal that the string's backreaction is well-approximated by superposition of gravitational shock waves, \GSW, rests on several important assumptions, which we will try to justify in turn.

\paragraph{Neglecting string tension:} 
Firstly, we require that the interaction between the individual string bits does not contribute appreciably to the backreaction of the string.  Admittedly such an assumption would fail for a slowly-moving string where longitudinal boost invariance of the string's stress tensor implies that the (negative) pressure is of the same magnitude as the energy density.  However, under transverse boosts with Lorentz factor  $\gamma_{\perp}$, energy density is enhanced  by a factor of $\gamma_{\perp}^2$ compared to the pressure which captures the effect of interactions.  In fact, this is not surprising, given the parton model of the string \cite{Kogut:1972di}. 
This means that our assumption of non-interaction becomes arbitrarily good for high enough string's transverse velocity, so that in the relativistic regime we can indeed treat the string as composed of relativistic particles.  

\paragraph{GSW from each string bit:}
The second assumption is that each string bit produces a backreaction in the form of GSW, which is valid as long as the string's transverse velocity is sufficiently relativistic.  As observed above, \req{transvgamma} guarantees
$V_{\perp}^2 \, \gamma_{\perp}^2 \gtrsim 1$ whenever either the quark velocity is sufficiently relativistic ($v \, \gamma \gtrsim 1$) or the string bit under consideration is sufficiently deep in the bulk ($u \gtrsim u_h$), which quantifies the regime of validity of 
our second assumption.  We note that this regime of validity happily coincides with our regime of interest: in particular, we are interested in $\CE$ at $r \gtrsim R_0/v$ where the notion of radiation is well defined, which indeed originates from the string bits at $u \gtrsim u_h$.

\paragraph{Linear superposition of GSWs:}
Finally, we assume that the individual GSWs superpose linearly, despite the nonlinearities of general relativity.  Although this may not hold for arbitrary GSWs -- for example head-on collisions can produce black holes (see e.g.\ \cite{Grumiller:2008va,Gubser:2008pc}, and more recent work reviewed in \cite{Kovchegov:2010zg}) -- it does hold in our case: examining the geometry in detail, we can show that the individual GSWs intersect mildly enough (in the sense indicated in the next section), and their interaction is strongest at the string itself.  Because we started with a nearly-test string, the GSWs produced by its backreaction will themselves backreact parametrically less (by ${\mathcal O}(1/N_c) \to 0$ in the present context).
Hence, given that  \cite{Athanasiou:2010pv} justifiably used linearized gravity to calculate the string's backreaction, it is certainly valid to use linearized gravity for the GSWs' backreaction.
This justifies superposing individual GSWs to estimate the backreaction of the string.

\subsection{Bulk construction of \GSW}
\label{s:GSWconstruct}

Having justified why we may take the string to backreact in a manner which is captured by \GSW, we now proceed to construct such a superposition explicitly.
The key step  is understanding how a single string bit, moving relativistically in the direction of $\transv$, backreacts on AdS.  Such a metric is given by the AdS analog of the Aichelburg-Sexl metric \cite{Aichelburg:1970dh} describing a gravitational shock wave in flat spacetime.\footnote{
 To write down the AdS analog of Aichelburg-Sexl metric describing a GSW in AdS, one can either boost the black hole metric as  in \cite{Horowitz:1999gf}, or one can glue two pieces of AdS along a `null plane' as discussed in \cite{Sfetsos:1994xa} and \cite{Dray:1984ha}; this was utilized for context of AdS$_4$ by \cite{Hotta:1992qy} and studied further in \cite{Podolsky:1997ni}.
} 
Geometrically, a GSW in flat spacetime is supported on a `transverse null plane'; in AdS, which is conformally flat, there is a corresponding transverse null hypersurface.

Although one could construct this explicitly by a similar procedure to \cite{Horowitz:1999gf}, we will instead use a trick, detailed in \App{a:GSWconstr}.  We observe that a spatial slice of GSW at a fixed time forms a hypersurface which is generated by spacelike geodesics, emanating from its source in a direction normal to the source's velocity $\transv$.  While this is readily apparent in the coordinates used in \cite{Horowitz:1999gf} simply by symmetry arguments, the statement as such is a geometric one, and therefore holds in any coordinate system.  This means that to find the support of a GSW in AdS at constant $t$, written in the form \req{AdSmet}, we only need to find the constant-$t$ 
spacelike geodesics, whose initial position $p_i$ lies along the string and initial velocity points normally to $\transv$.

As we show explicitly in  \App{a:GSWconstr}, such geodesics form a `hemisphere' in the $(x,y,z,u)$ space, whose equator lies on the boundary $u=0$ and whose radius and position is determined by the geometry of the string.  The `intersection' of the GSW with the AdS boundary determines the spatial support of the boundary energy density $\CE$ at a fixed time, which forms a sphere in $(x,y,z)$ space and therefore a circle on the orbital $(x,y)$ plane.  Apart from knowing the support of $\CE$ sourced by a given string bit, we are interested in knowing its magnitude.  This will be discussed in \sec{s:bdystress}, but as a zeroth order estimate, we can determine where  $\CE$ dominates.  At each bulk radius $u$, the dominant contribution of the GSW lies along the direction nearest to the source; geometrically this represents a `steepest' bulk spacelike geodesic.  The endpoint $p_0$ of this geodesic singles out a position at which $\CE$ peaks.

\begin{figure}
\begin{center}
\includegraphics[width=3.5in]{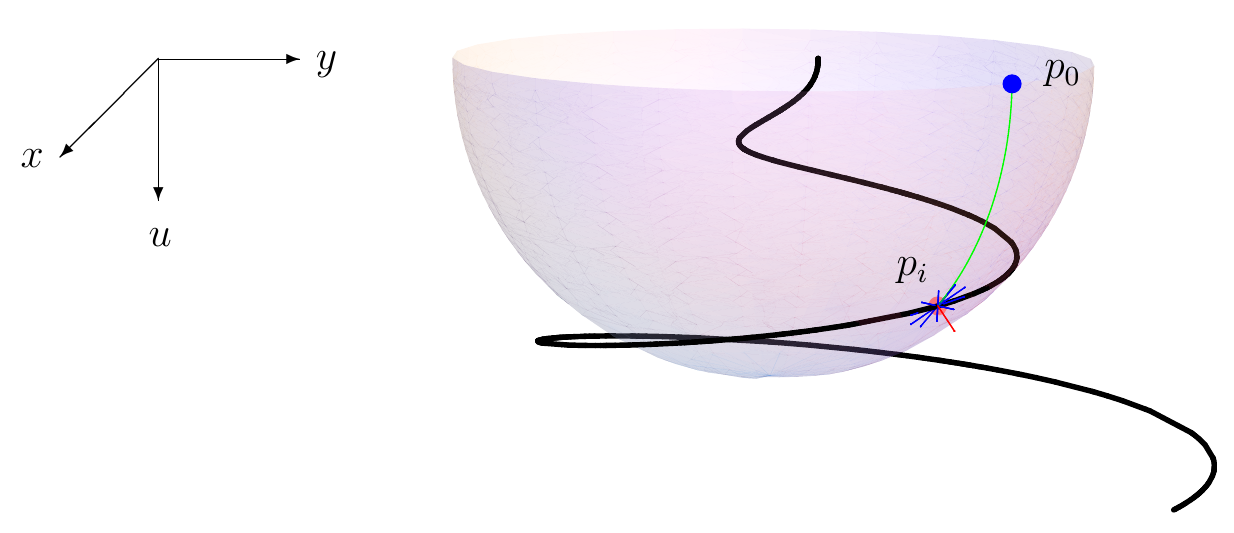}
\hspace{-1cm}
\includegraphics[width=3.2in]{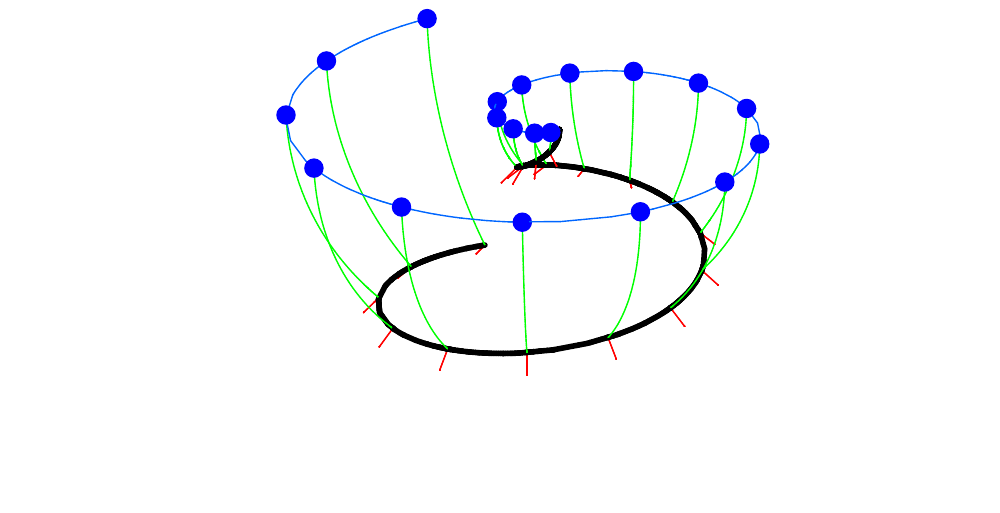}
\caption{
{\bf Left:} Construction of the GSW sourced by a given string bit: The thick black line is the string (corresponding to $R_0 = 1, v = 0.7$). We pick an initial point $p_i$ (here we chose $u_i=7$).  We then construct the transverse velocity (red line segment) and consider all geodesics emanating perpendicularly to it (blue line segments).  This generates the full GSW (surface), with the shortest geodesic to the boundary (green curve) ending on the (blue point) $p_0$. 
{\bf Right:} The dominant contribution  $p_0$ from a set of representative points along the bulk string.
We can see the beginnings of the boundary spiral along which expect $\CE$ to peak (blue curve).
}
\label{f:GSW_constr}
\end{center}
\end{figure}

The construction indicated above and detailed in \App{a:GSWconstr} is illustrated in \fig{f:GSW_constr} (left),  which shows a single GSW sourced by the string bit at $p_i$, with the green geodesic and its endpoint $p_0$ corresponding to the dominant contribution of the GSW.
So far, we have indicated how to construct the GSW sourced by a particular string bit, but we can simply repeat the construction for all string bits, as indicated in the right panel of \fig{f:GSW_constr} (which is further tilted for greater ease of visualization) for a representative set of GSWs.  
We see that the (blue) points $p_0(u_i)$ on the boundary, indicating the dominant contribution to $\CE$, indeed lie on a spiral, as calculated in \cite{Athanasiou:2010pv}.  In \sec{s:results} we will examine this spiral in greater detail.

\subsection{Boundary energy density induced by \GSW}
\label{s:bdystress}

Now that we have determined the bulk locus of a given gravitational shock wave, we can trivially extract the {\it support} of the boundary stress tensor induced by such a metric deformation.  
The essential tool from AdS/CFT is that the asymptotic form of the bulk metric deformation $\delta g_{\mu\nu}$ away from AdS determines the boundary stress tensor expectation value $\expval{T_{\mu\nu}}$ as we review in \App{a:stress} (cf.\ equations \req{genmetdef} and \req{Tfromgenmetdef}).
In particular, we know that 
$\expval{T_{\mu\nu}} \ne 0$ only where the GSW intersects the boundary.  
However, to determine the value of $\expval{T_{\mu\nu}}$ at this locus, we need to make additional use of the GSW's {\it profile}.  Here we describe a way to estimate this.

The proper calculation would be to generate the explicit GSW metric corresponding to a source propagating along some arbitrary null direction in AdS, for instance by boosting a Schwarzschild-AdS black hole to speed of light keeping the total energy fixed.  Once obtained, we could calculate the boundary stress tensor induced by this metric, either directly via a Brown-York type procedure using the covariant form of \cite{Balasubramanian:1999re},
or by first recasting the metric into the Fefferman-Graham form by a suitable change of coordinates and then reading off the stress tensor as in \cite{deHaro:2000xn}.
Instead, here we take a shortcut, by using previous calculations \cite{Gubser:2008pc} as described in \App{a:stress}.  In particular, we recast the induced stress tensor on the boundary in terms of geometrical properties of the geodesics described above.

To write this compactly, consider a GSW at a given time $t$, sourced at the bulk point $p_i$.  The boundary light-up (i.e.\ induced energy density $\CE$) occurs at all points $p_f$ reachable from $p_i$ by a spacelike geodesic, normal to the source's velocity.  Let the regularized proper length\footnote{
Since the proper length along a spacelike geodesic between any bulk point and a point on the boundary is infinite, we need to regulate this quantity.  However there is a standard procedure of doing this, as mentioned in \App{a:GSWconstr} which we employ.
} along such geodesics be denoted by $\CL(p_i, p_f)$.
In \App{a:stress} we argue that the profile for the boundary energy density $\CE(p_f)$ is given by
\begin{equation}
{\cal E}(p_f) \propto 
 e^{-3 \, \CL(p_i, p_f)} \ .
\label{EGSWprofile}
\end{equation}	
We will use this expression at the end of \sec{s:numer} below.  We see that the profile of $\CE$ peaks sharply at a point $p_0$ on the GSW located closest to the source.  For example, for a source moving  
parallel to AdS boundary along a trajectory $x=t$, $y=z=0$, and $u=u_i$, the induced energy density at time $t$ is
\begin{equation}
{\cal E}(t,x,y,z) 
\propto 
\frac{\delta(t-x) \, u_i^3}{\left(u_i^2 + y^2 + z^2 \right)^3 }  \ .
\label{EGSWprofilepar}
\end{equation}	
Once we have expressed the energy density due to a single string bit at $u=u_i$, we integrate this over all values of $u \gtrsim u_h$.  In fact, due to the transverse localization of each GSW, only a small range of $u$ contributes to the energy density in given small neighbourhood.

\section{Results}
\label{s:results}

In the preceding section we have explained how to obtain the boundary energy density $\CE$ sourced by \GSW\ corresponding to a relativistic string in the bulk.  Here we carry out this construction in detail for the circling quark discussed in \sec{s:setup}, and compare with the results of \cite{Athanasiou:2010pv}.  
We will first present our results graphically to confirm the qualitative features of $\CE$.  Although one could in principle obtain the corresponding expressions analytically, they would be far too lengthy and unilluminating to write explicitly.  However, since we are mainly interested in the behavior of $\CE$ in the asymptotic region (where the radiation is cleanly separated from the background), we can simplify these expressions by focusing on the $r \gg R_0$ regime.  Such asymptotic analysis will be carried out in \sec{s:asymp}, where we will extract the key  features  of $\CE$ quantitatively.
Since the time dependence is given by rigid rotation, it suffices to consider $\CE$ at $t=0$.

\subsection{Numerical results}
\label{s:numer}

Rather than estimating $\CE(t,r,\theta,\phi)$ using the full profile \req{EGSWprofile}, it is instructive to separate the analysis of $\CE$ into several parts.   We will first consider the dominant contribution to $\CE$ which will confirm the spiral shape and allow us to extract the spiral arm spacing.  We will then consider the full support of $\CE$, which will reveal the $v$-dependence of the width of the spiral arm.  We will also briefly consider the spiral peak profile.   Finally, we will examine the polar angle dependence of $\CE(\theta)$.

\paragraph{Spiral spacing:}
Let us first consider the boundary curve where $\CE$ is dominant, corresponding to the spiral indicated in the right panel of \fig{f:GSW_constr}.  The bulk counterpart is given by maximizing the metric deformation $\delta g_{\mu \nu}$ at each $u$. 
\begin{figure}
\begin{center}
\includegraphics[width=2.8in]{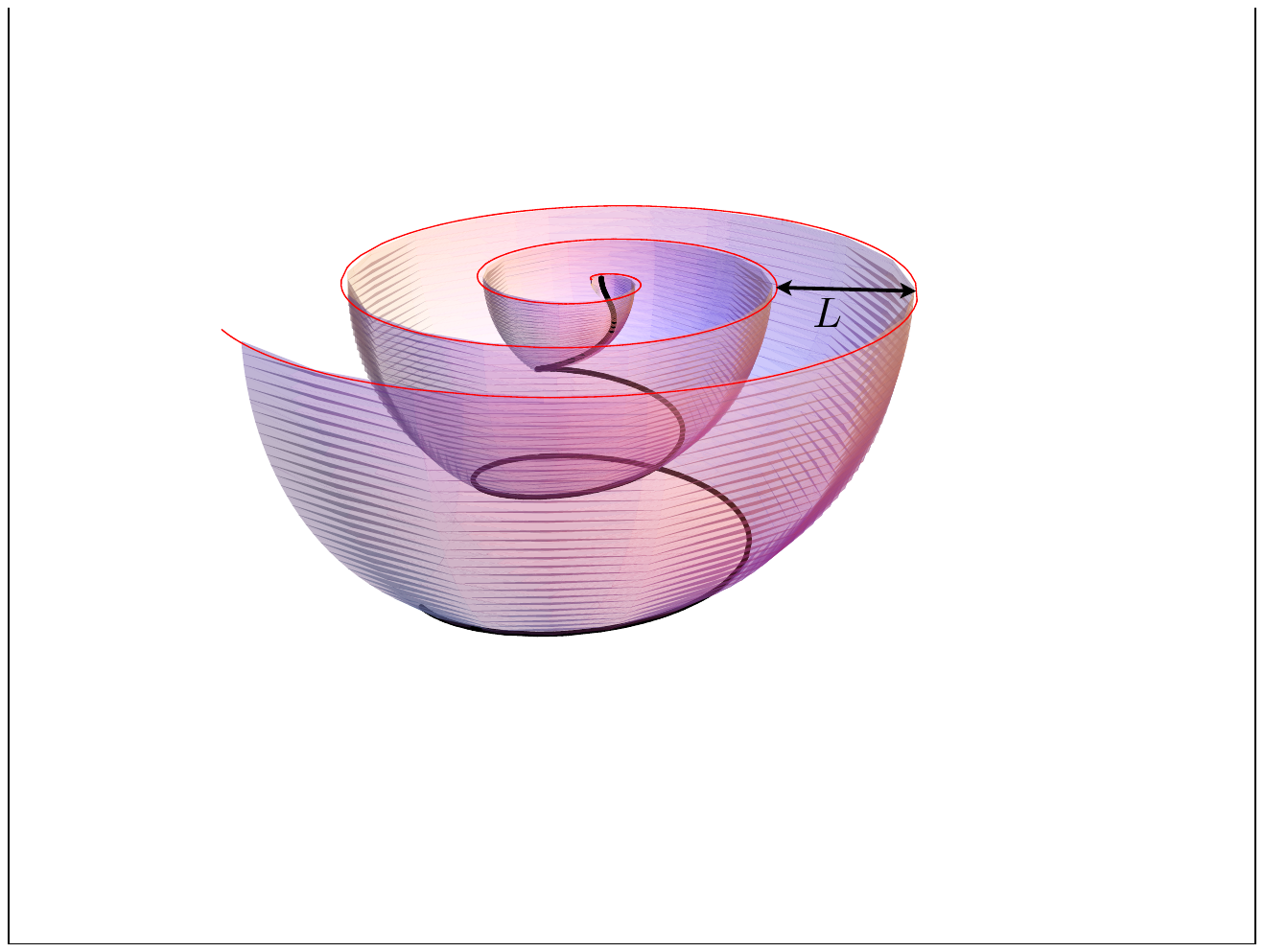}
\hspace{1cm}
\includegraphics[width= 2.5in]{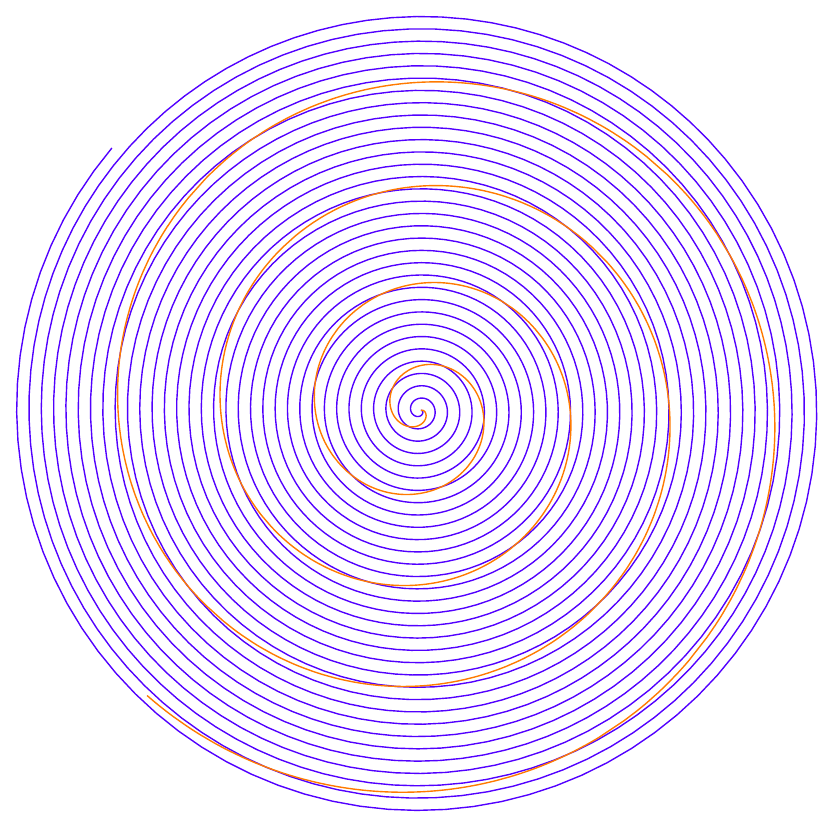}
\caption{
Dominant contribution from the \GSW, in bulk (left) and on boundary (right).
{\bf Left:}
The thick (black) curve denotes the string (induced by quark with parameters $v=0.5$ and $R_0 =1$). The surface plotted is the dominant part of \GSW, whose intersection with the boundary gives the expected spiral curve (the thin red curve at the top is its approximation \req{GSWlightup} discussed in \sec{s:asymp} below). 
{\bf Right:}
Spiral induced on boundary for two representative velocities: $v=0.1$ (orange) and $v=0.9$ (purple).}
\label{f:bulk_bdy_dom}
\end{center}
\end{figure}
For each string bit, parameterized by $u_i$, this occurs along the steepest geodesic (indicated by the green curve in \fig{f:GSW_constr}).  The family of the dominant parts of GSW($u_i$) is plotted in  \fig{f:bulk_bdy_dom} (left).  Its intersection with the boundary $u=0$ (top of the plot) generates the spiral formed by $p_0(u_i)$.  Such a spiral is plotted in \fig{f:bulk_bdy_dom} (right) for two distinct velocities, $v=0.1$ and $v=0.9$.  Both spirals have qualitatively correct behaviour, with the spiral arms growing linearly with angle, but this rate of growth scales inversely with $v$.  In particular, we find that the distance between the spiral arms for general parameters $R_0$ and $v$ is given by 
$L \approx \frac{2 \, \pi \, R_0}{v}$, in agreement with \cite{Athanasiou:2010pv}.  This will be confirmed analytically in \sec{s:asymp}.
Note that periodicity of the quark's motion implies that the spiral rotates rigidly with periodicity $T = \frac{2 \, \pi }{\omega_0} =\frac{2 \, \pi \, R_0}{v} =L$, which in turn implies that the spiral arms move outward at the speed of light, independently of the velocity $v$.  
This provides the most crucial check of the beaming mechanism.

\paragraph{Spiral width:}
So far we have confirmed that the dominant contribution to the boundary energy density induced by \GSW\ indeed lies on a spiral with correct spacing of spiral arms and time dependence. 
In order to extract further features of the spiral, such as its width, we need to consider more than just the dominant contribution of each GSW.
Let us therefore focus on the full support of the GSW, but instead ignore the variations of its intensity.
In other words, the boundary light-up sourced by each string bit is a full sphere rather than a single point.  On the equatorial $(x,y)$ plane of the boundary such light-up is a circle, and the superposition of all circles indicates of the support of $\CE(x,y)$ due to the entire string.

\begin{figure}
\begin{center}
\includegraphics[height=2.8in]{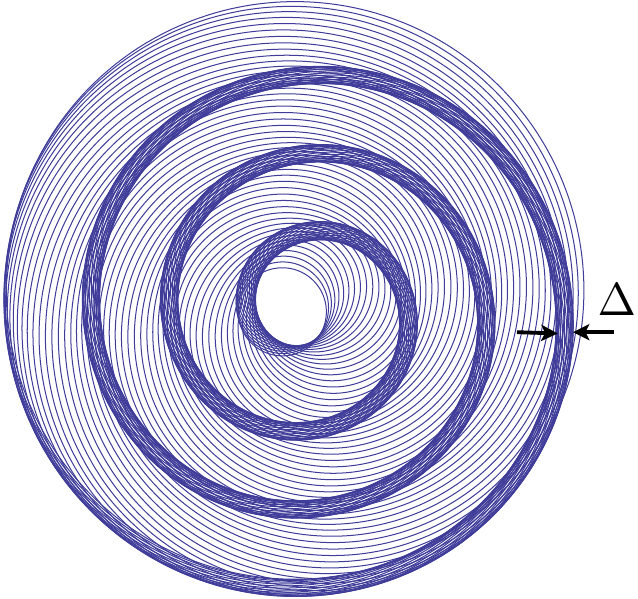}
\hspace{1cm}
\includegraphics[height= 2.8in]{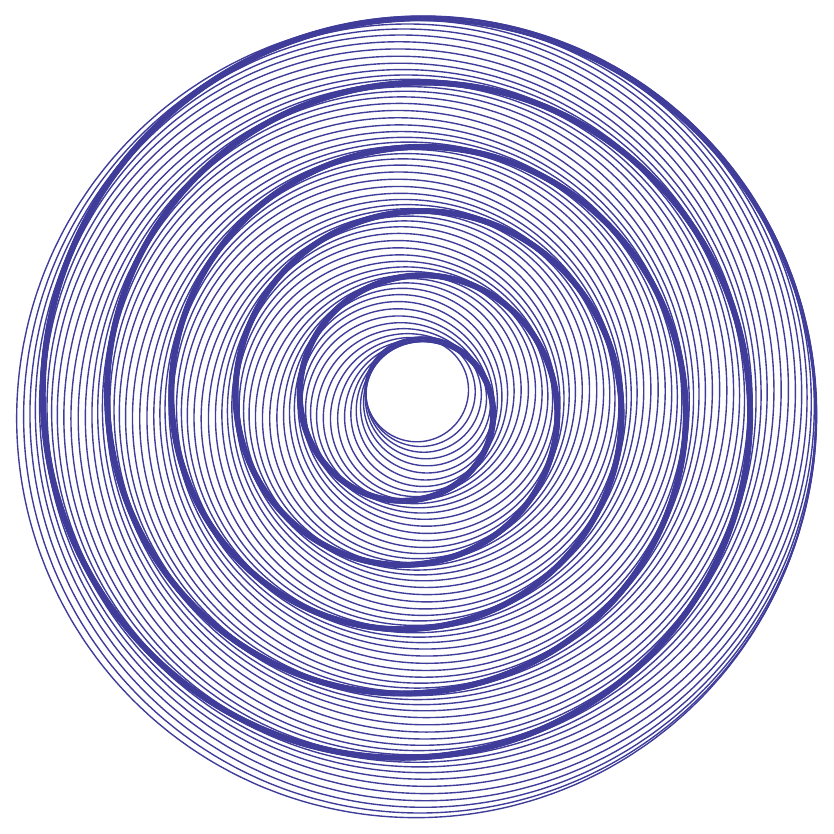}
\caption{
Light-up at $t=0$, $\theta = \frac{\pi}{2}$, and $u=0$ from GSWs originating at various points along the string (corresponding to quark velocity $v=0.5$  {\bf (left)} and $v=0.7$ {\bf (right)} and $u\in(5,40)$ in increments of 0.3).  Each GSW forms a circle approximated by \req{singleGSW}; but from the superposition of all such circles, there clearly emerges an overall spiral pattern which matches that of \fig{f:bulk_bdy_dom}, approximated by \req{GSWlightup}.
}
\label{f:gswsuperpos}
\end{center}
\end{figure}

Figure \ref{f:gswsuperpos} shows the boundary light-up from \GSW, for quark velocity $v=0.5$ (left) and $v=0.7$ (right).   While only the individual GSWs are plotted
(each one a circle with radius and origin approximated by \req{singleGSW}, as discussed in \sec{s:asymp}), we see that an overall spiral pattern indeed emerges from the superposition of such circles.  This spiral of course agrees with that seen in \fig{f:bulk_bdy_dom} (approximately given by \req{GSWlightup}), but now we can also see that it has a sharp width $\Delta$, which gets smaller with increasing $v$.  
Numerically we confirm agreement with the scaling $\Delta \sim \gamma^{-3}$ found in \cite{Athanasiou:2010pv} and familiar in weakly-coupled contexts; in \sec{s:asymp} we will be able to go further in extracting both this scaling and its coefficient.

The most striking feature of Figure \ref{f:gswsuperpos} is that the spiral width (corresponding to the energy pulse discussed in Section \ref{s:setup}) does {\it not} broaden with radius $r$, but rather remains constant along the full spiral, despite the fact that the radius of a GSW sourced at $u_i$ grows with $u_i$.  
Although a-priori one might worry that this feature is an outcome of considering the full support of the GSWs, it is nevertheless possible to check that the spiral peak remains of constant width even with the actual profile of each GSW taken into account.

\begin{figure}
\begin{center}
\includegraphics[width=2.5in]{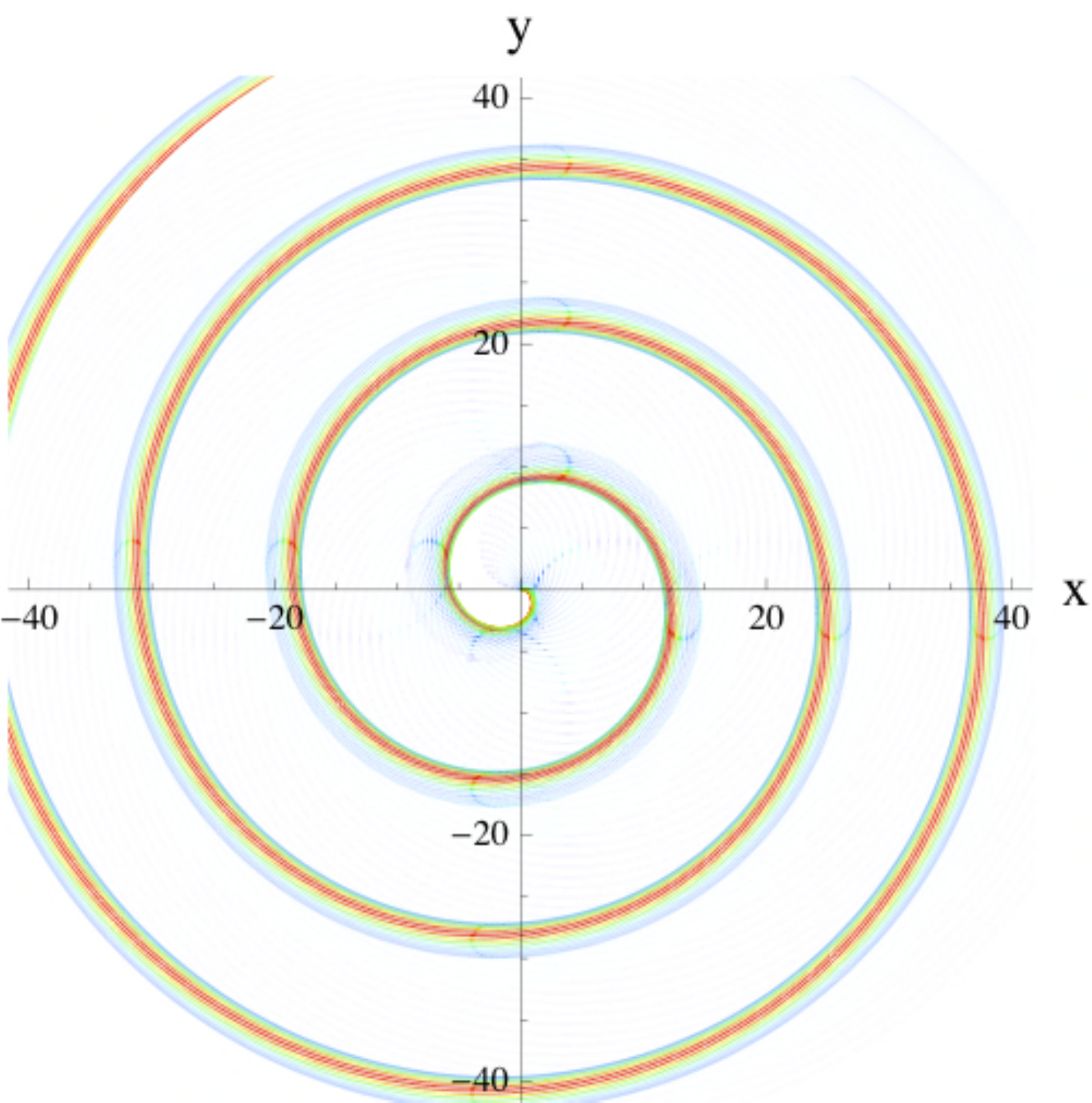}
\caption{Indication of the spiral peak profile.  The plot is as in left panel of \fig{f:gswsuperpos}, but we have added color-coding and opacity to reflect the strength of the GSW on the boundary (red indicates higher energy, blue lower energy).
}
\label{f:spiral_peak_col}
\end{center}
\end{figure}

\paragraph{Spiral peak profile:}
We can now combine the calculations summarized in \fig{f:bulk_bdy_dom} and \fig{f:gswsuperpos}, and consider the superposition \GSW, by taking into account the 
actual profile \req{EGSWprofile}.   Since $\CE(r)$ falls off with $r$ (due to the corresponding source  being further in the bulk), we can instead consider the rescaled energy density $r^2 \, \CE$ as done in \cite{Athanasiou:2010pv}. 
While conceptually straightforward, the determination of $\CE(r)$ is still computationally rather involved, so we will resort to graphical means to estimate the profile. Specifically, in \fig{f:spiral_peak_col} we use color-coding and opacity (modulated by the profile \req{EGSWprofilepar}) to indicate how the full superposition  behaves.
We see that the resulting profile of $\CE$ is again in qualitative agreement with that calculated in \cite{Athanasiou:2010pv}.  It would be interesting to examine the peak profile more closely to see how precisely we can recover the quantitative shape $\CE(r)$ computed in \cite{Athanasiou:2010pv}.  In particular, the latter has a curious feature of $\CE$ becoming slightly negative at the tails of the peak.\footnote{
Although the profile \req{EGSWprofile} is manifestly positive, we have not separated out the energy density due to the quark's field.  It is conceivable, though remains to be checked, that once this is properly taken into account, the slightly negative values could be reproduced.}

\begin{figure}
\begin{center}
\includegraphics[width=2in]{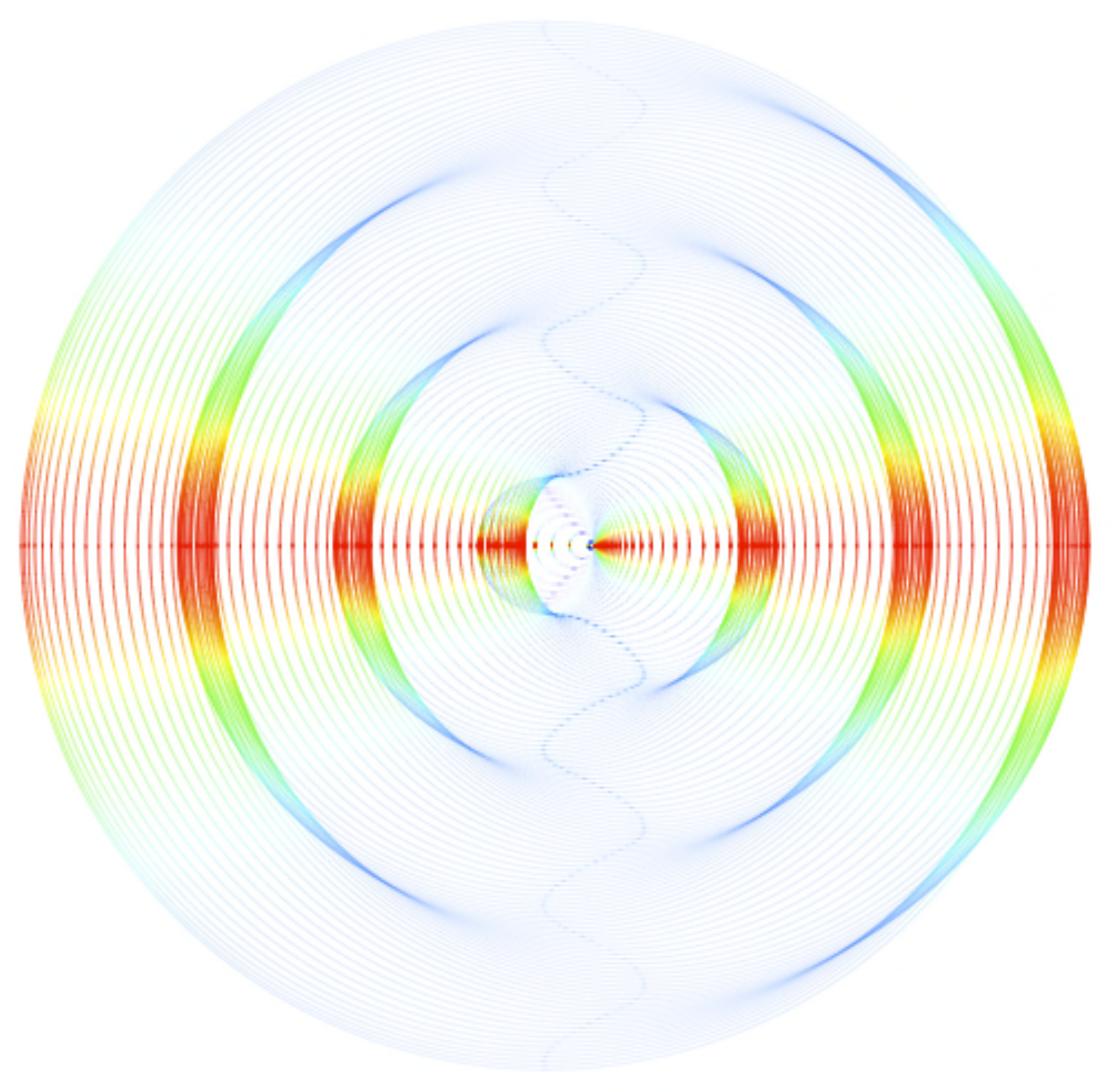}
\hspace{1.5cm}
\includegraphics[width =2.5in]{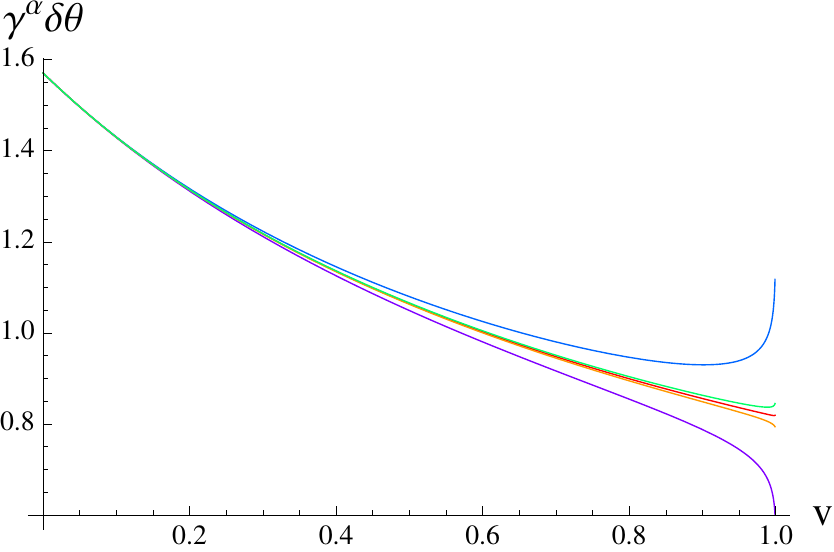}
\caption{Dependence of $\CE$ on the polar angle $\theta$.
{\bf Left:} \GSW\ light-up $\CE(x,z)$ for $v=0.5$; color-coding and opacity indicate the relative fall-off in $\theta$.
{\bf Right:} Plot of $\gamma^{\alpha} \, \delta \theta$ as a function of the quark velocity $v$ for various values of $\alpha$; from top to bottom, $\alpha = 1.1$ (blue), $1.01$ (green), $1$ (red), $0.99$ (orange), and $0.9$ (purple).  This indicates that $\delta \theta \sim \frac{1}{\gamma}$.
}
\label{f:delta_theta}
\end{center}
\end{figure}

\paragraph{Dependence of $\CE$ on the polar angle:}
We have hitherto considered the energy density $\CE(x,y)$ in the orbital plane, but we can use the same methods to find the full $\CE(x,y,z)$ (to be compared with the  cutaway plot presented in \cite{Athanasiou:2010pv}'s Fig.4).
Due to the orbital symmetry of the set-up, it suffices to focus on the $(x,z)$ plane.
This is presented in \fig{f:delta_theta} (left); as in \fig{f:spiral_peak_col}, we plot the full support of a representative set of GSWs, with color and opacity modulated by the profile $\CE(\theta)$ (though for simplicity we haven't folded in the radial dependence).
In particular, we find that the GSWs superpose along crescent-shaped regions, for which $\CE(\theta)$ decreases off the orbital plane (horizontal axis) as indicated, with characteristic width $\delta \theta$, which we approximate by the angular span of these crescents.  In the right panel of  \fig{f:delta_theta}, we analyse the dependence of this width $\delta \theta$ on the quark velocity $v$, by plotting $\gamma^{\alpha} \, \delta \theta$ as a function of $v$ for several values of the exponent $\alpha$.  Although for our definition of $\delta \theta$ none of these curves is a constant, there is a clear indication that as $v \to 1$, only the exponent $\alpha = 1$ gives an approximately constant behavior.  This confirms the result of \cite{Athanasiou:2010pv} that
the characteristic spread $\delta \theta$ decreases with increasing $v$ as $\Delta \theta \sim 1 / \gamma$.

\subsection{Asymptotic analysis}
\label{s:asymp}

Although the results presented above in a graphical format can be obtained analytically, the exact expressions are far too lengthy and unilluminating.
However, we are not so interested in the exact answer of $\CE(t,r,\theta,\phi)$ for the full range of $r$, as in a more restricted asymptotic regime $r \gg R_0$ where the notion of radiation can be easily disentagled from the quark.  
Correspondingly, in the bulk we likewise wish to focus on the region $v \, \gamma \, u \gg R_0$, which is simultaneously the region where the \GSW\ approximation is most accurate.
This allows us to simplify the exact expressions for the string embedding \req{string}, and approximate\footnote{
Note that it is important to keep the first subleading term in $\phi(u)$ (which originates from the $\tan^{-1}( - \frac{ v \, \gamma \,  u }{R_0})$ term in  \req{string}), since it introduces a phase shift in the quantities of interest and therefore is visible even in the asymptotic regime.
}
 the full string by a linearly-flaring-out helix,
\begin{equation}
R(u) \approx  v \, \gamma \,  u  \qquad {\rm and} \qquad
\phi(u) \approx \frac{\pi}{2} - \frac{ v \, \gamma \,  u }{R_0} \ .
\label{stringasym}
\end{equation}	
That in turn enables us to obtain compact closed-form expressions for the boundary light-up due to the resulting \GSW.
In this subsection, we summarize such analysis, which complements the numerical results of \sec{s:numer}.  

The approximation of the string profile \req{stringasym}
 simplifies the transverse velocity \req{transv}, from which we can extract the constants of motion describing the steepest geodesic plotted in \fig{f:GSW_constr}.
Having the constants of motion and the initial point, we can now construct the full geodesic, and thereby find the coordinates of its deepest point, 
$p_m = (0,x_m,y_m,z_m,u_m)$, as well as the dominant light-up $p_0 = (0,x_0,y_0,z_0,0)$, using \req{geodprof}.  
The boundary projection of $p_m$ gives the origin of the boundary light-up due to string bit at $u$ (for compactness of notation, we will drop the subscript from $u_i$), whereas $p_0$ corresponds to the peak of this light-up.
Specifically, for the string bit at $u$, 
\begin{equation}
x_m
\approx \frac{R_0 \, s}{v^2} \ , \qquad
y_m \approx -\frac{R_0 \, c}{v^2} \ ,
\qquad 
z_m = 0 \ ,
\label{pmas}
\end{equation}	
\begin{equation}
x_0 \approx  \frac{R_0 \, s}{v^2}  + 
\sqrt{\frac{v^4 \, \gamma^2 \, u^2 +R_0^2}{v^6 \, \gamma^2 \, u^2 +R_0^2}} 
\, \left( v \, \gamma \, u \, c  - \frac{R_0 \, s}{v^2} \right) 
\, , \ \
y_0 \approx  - \frac{R_0 \, c}{v^2}  + 
\sqrt{\frac{v^4 \, \gamma^2 \, u^2 +R_0^2}{v^6 \, \gamma^2 \, u^2 +R_0^2}} 
\, \left( v \, \gamma \, u \, s  + \frac{R_0 \, c}{v^2} \right) \ ,
\label{xy0as}
\end{equation}	
and $z_0 = 0$,
where 
\begin{equation}
c(u) \equiv \cos \left(\frac{\pi}{2} - \frac{ v \, \gamma \,  u }{R_0} \right)
\ands
s(u) \equiv \sin \left(\frac{\pi}{2} - \frac{ v \, \gamma \,  u }{R_0} \right) \ .
\label{}
\end{equation}	
Note that the origin of the GSW from a point $u$ along the string describes a circle of constant radius $\frac{R_0}{v^2} $ in the $(x,y)$ plane on the boundary, with angular dependence parameterized by $u$. 

To summarize, the boundary light-up from a GSW sourced by a string bit at $u$ is supported on a sphere of radius
$\rho_m$ with origin  at $(r_m, \theta_m, \varphi_m)$, 
 given approximately by 
\begin{equation}
\rho_m \approx \sqrt{\gamma^2 \, u^2 +\frac{R_0^2}{v^4}} \ ; \qquad
r_m \approx \frac{R_0}{v^2} \ , \ \ 
 \theta_m = \frac{\pi}{2}\ , \ \ 
\varphi_m \approx - \frac{v \, \gamma \, u}{R_0} \ .
\label{singleGSW}
\end{equation}	
The amplitude of this GSW is strongest nearest the source; within the boundary, the dominant light-up due to the string bit at $u$ occurs at 
\begin{equation}
r_0 \approx \gamma \, u \ , \ \ 
\theta_0 = \frac{\pi}{2} \ , \ \ 
{\rm and} \ \
\varphi_0 \approx \frac{\pi}{2} - \frac{v \, \gamma \, u}{R_0} \ .
\label{GSWlightup}
\end{equation}	
Note that this indeed gives a spiral curve on the orbital plane, parameterized by $u$, whose radius $r$ grows linearly with angle $\varphi$.

The spacing $L$ between the spiral arms (cf.\ \fig{f:bulk_bdy_dom}) can now be obtained by examining how $r_0$ increases under a single orbit $\Delta \varphi = 2 \pi$: $u$ jumps by $\Delta u = \frac{2 \pi \, R_0}{v \, \gamma}$, which immediately gives the previously quoted result
\begin{equation}
L =\frac{2 \, \pi \, R_0}{v} \ .
\label{}
\end{equation}	
Apart from verifying the expected spiral spacing and therefore the speed of propagation of radiation bursts, we can now also confirm the phase, which likewise agrees precisely with the results of \cite{Athanasiou:2010pv} and the weakly coupled synchrotron radiation.

This width $\Delta$ of the spiral arm (cf.\ \fig{f:gswsuperpos}) is rather more involved, albeit conceptually straightforward to extract.
The calculation proceeds by finding intersections of `adjoining' GSWs, at $u$ and $u+\varepsilon$ in the limit as $\varepsilon \to 0 $ (there are two solutions for each circle, and under variation of $u$ they form two nested spirals).  These give the edges of the emergent spiral in \fig{f:gswsuperpos}.  We determine its width $\Delta$ by taking the difference of the edges' intersections with a given radial ray.  This yields
\begin{equation}
\Delta = \frac{2 \, R_0}{v} \, \left(\frac{1}{v \, \gamma}  - \tan^{-1} \frac{1}{v \, \gamma}  \right) \ .
\label{}
\end{equation}	
While this expression applies for any $v$ in the asymptotic regime $u \gg \frac{R_0}{v^2}$, it is more interesting to consider the scaling of the thickness $\Delta$ at large velocities $v \sim 1$, i.e.\ as $\gamma \to \infty$.  In this limit we find that
\begin{equation}
\Delta \to \frac{2 \, R_0}{3 \, v^4} \, \frac{1}{\gamma^3}  \qquad {\rm as} \qquad v \to 1 \ . 
\label{}
\end{equation}	
This confirms the expected scaling $ \Delta \sim \gamma^{-3}$, which was found in  \cite{Athanasiou:2010pv}.  In fact, there the authors conjectured that the natural length scale would be given by 
$ \Delta \sim \frac{ R_0}{\gamma^{3} \, v^2}$; here we confirm the $R_0$ scaling and the coefficient up to a $\frac{2}{3v^2} \sim \CO(1)$ factor.

\section{Discussion}
\label{s:extensions}

In the preceding section we have described the important features of the boundary energy density $\CE(t,r,\theta, \varphi)$ induced by a relativistic coiling string in the bulk.  We proposed approximating this by a superposition of gravitational shock waves produced by each string bit -- we dubbed this the `beaming mechanism' -- and calculated the consequences of this proposal.  In particular, we have obtained the characteristic spiral pattern of synchrotron radiation (cf.\ e.g.\ \fig{f:spiral_peak_col}).  We have calculated the spacing, phase, width, speed of propagation, and polar extent, of each energy pulse, and found each of these characteristic features in complete quantitative agreement with the computations of \cite{Athanasiou:2010pv}.

Our results suggest that it is indeed valid to view the string's backreaction as given by superposition of gravitational shock waves.  
From a computational perspective, this presents a major simplification:  instead of having to solve the full linearized Einstein's equations in the bulk, we merely need to find geodesics.  In fact, this provides a useful tool for understanding the radiation of a relativistic quark following an arbitrary trajectory, since the string tends to become more relativistic deeper in the bulk and develops a worldsheet horizon, making the \GSW\ approximation more accurate further away from the quark.  Since for any quark trajectory the corresponding string motion is known explicitly \cite{Mikhailov:2003er}, we can construct the \GSW\ just as easily as for the synchrotron case discussed above.

Encouraged by the success in using \GSW\ in AdS to reproduce the synchrotron radiation in strongly coupled medium at zero temperature calculated in \cite{Athanasiou:2010pv}, a natural question to ask is whether this method also applies in the case of non-zero temperature.  
Indeed, once one generalizes the GSW construction to Schwarzschild-AdS background, one should likewise be able to estimate at least some part of the quark's radiation in a thermal strongly-coupled plasma, which should suffice to reproduce, for instance, the diffusion wake and sonic boom generated by a moving quark \cite{Friess:2006fk}.  However, it is important to realize that in the presence of a bulk AdS black hole, 
the string no longer remains relativistic deep in the bulk, so the \GSW\ approximation will have much more constrained regime of validity -- in particular, it will no longer improve with distance from the quark.   Zeroth order expectations from directly adapting our geodesic construction would suggest that the radiation which starts out collimated diffuses on a thermal scale; as is indeed consistent with expectations form the field theory side.  

Moreover, the construction of a gravitational shock wave itself, in Schwarzschild-AdS background, is substantially more involved and is it no longer guaranteed that the metric perturbation from Schwarzschild-AdS due to a relativistic particle would remain similarly localized.\footnote{We thank Gary Gibbons for pointing out this contingency.}
Nevertheless, given some region where the \GSW\ approximation is accurate, it may be possible to go beyond the initial regime of validity by performing  series expansion in $1/\gamma$ and building up the backreaction effects order-by-order.  We however leave this for future work.

Finally, we end with the issue which originally motivated the present work, namely how the observed beaming effect bears on the scale/radius duality.
Recall that  scale/radius (or UV/IR) duality \cite{Susskind:1998dq}, which has been one of the first and most important entries in the AdS/CFT dictionary, suggests that the closer a bulk excitation is to the AdS boundary, the smaller is the characteristic size of the dual CFT excitation.  This provides an important insight into the holographic nature of the AdS/CFT duality: the `extra' (radial) dimension 
in the bulk arises from a scale in the dual CFT.  This intuition has
bolstered our understanding of 
various dynamical processes.  For example, the field theory notion of color transparency, that different-scale excitations pass through each other without interacting, is beautifully explained by bulk locality: in the bulk, the corresponding excitations do not interact due to being separated in the radial direction.  From this, one might have naively expected that same-scale excitations passing through each other would interact.

However, the present context emphasized that naive applications of the scale/radius duality should be treated with caution.  
In particular, as we have seen, a source deep in the bulk of AdS may nevertheless produce a sharply-localized effect on the boundary.  By taking an ordinary source and a beamed source, we can engineer a situation where two excitations with (instantaneously) similar characteristic scale do not interact (because within the bulk they are well separated), as well the converse situation where two excitations with widely differing scale nevertheless do interact (because the dual bulk excitations collide with each other).
The beaming mechanism, which underlies such constructions, is quite interesting.
The relativistic velocity of the string causes the backreaction due to each string bit to be localized on a co-dimension one surface of the GSW.  Although the size of this surface grows with the bulk position of the source (it has larger radius for the string bits deeper in bulk), the dominant effect is sharply peaked.  The final ingredient, however, is the superposition of all the GSWs from individual string bits, as emphasized most clearly by \fig{f:gswsuperpos}.  It is due to this combined effect that the radiation pulse on the boundary does not broaden as it propagates outward.

\subsection*{Acknowledgements}
\label{acks}

It is a pleasure to thank
Paul Chesler,
Roberto Emparan,
Gary Gibbons,
Gary Horowitz,
Hong Liu,
Shiraz Minwalla,
Rob Myers,
Joe Polchinski,
and Mukund Rangamani
for valuable discussions.
I would also like to thank MIT, UBC, Galileo Galilei Institute, and Imperial College London for their hospitality during this project.
This work is supported in part by a STFC Rolling Grant.

{\small 
\appendix
\section{Construction of GSW via geodesics}
\label{a:GSWconstr}

In this Appendix, we describe how to obtain the metric due to a gravitational shock wave in AdS, and a simplification for constructing the support of the GSW using geodesics.  

\paragraph{Defining GSW:}
Although one could construct a GSW in AdS directly in a manner similar to \cite{Horowitz:1999gf},
we will instead simply use their results, adapted to the present context.  
The method employed by \cite{Horowitz:1999gf} involves boosting an AdS black hole to the speed of light while keeping the total energy fixed.  Conveniently, they use coordinates in which the AdS$_d$ metric is manifestly conformally flat and Lorentz invariant (though not static),
\begin{equation}
ds_{\rm AdS}^2 = \frac{4 \, \eta_{\mu\nu} \, dy^\mu \,  dy^\nu}{(1 - \eta_{\alpha\beta} \, y^\alpha \, y^\beta )^2} \ ,
\label{LorAdS}
\end{equation}	
and utilize the known the shock wave solution in Minkowski space.
Without loss of generality, one can consider a massless particle moving in the $-y_1$ direction, 
and write an ansatz for its backreaction as
\begin{equation}
ds^2 = ds_{\rm AdS}^2 + \delta(y_+) \, \frac{f(\rho)}{(1+ y_+ \, y_- - \rho^2)} \, dy_+^2
\qquad {\rm where} \qquad
y_{\pm} \equiv y_0 \pm y_1 \ , \ \  \rho^2 = \sum_{i=2}^{d-1} y_i^2 \ .
\label{shockwavemet}
\end{equation}	
The function $f(\rho)$ can then be obtained by solving Einstein's equations for this ansatz.\footnote{
These take the simple linear form in terms of the Laplacian $D^2$ on the transverse space 
$ds_{\perp}^2 = \frac{d\rho^2 + \rho^2 \, d\Omega^2}{(1-\rho^2)^2}$:
$D^2 \, f - 4 \, (d-2) \, f \propto \delta(\rho)$, 
whose behavior near $\rho=0$ mimics that in flat spacetime.}
However, we first consider the {\it support} of the GSW, for which the exact form of $f$ is immaterial, and only the $\delta(y_+)$ term is relevant (we will return to considering the profile in \App{a:stress}, where we will start with a slightly different form of the metric).

To make contact with our set-up, we wish to recast this into the Fefferman-Graham form \req{AdSmet}.
Although a change of coordinates of the form 
\begin{equation}
t=  \frac{- 1+{\tilde y}_- \, {\tilde y}_+ - {\tilde \rho}^2}{2 \, {\tilde y}_-} \ , \qquad
r=  \frac {{\tilde \rho}}{{\tilde y}_-}\ , \ \
\cos \theta = -\frac{{\tilde y}_{4}}{{\tilde \rho}} , \ \
\tan \phi = \frac {{\tilde y}_{3}}{{\tilde y}_2} \ , \qquad 
u= - \frac{1+{\tilde y}_- \, {\tilde y}_+ - {\tilde \rho}^2}{2 \, {\tilde y}_-} \
\label{LorPoincCX}
\end{equation}	
considered by \cite{Horowitz:1999gf} corresponds to a radial null geodesic 
$u=t$ at $r= 0$, 
we could generalize this to arbitrary direction, by simply making an extra change of the $y$ coordinates, $y^{\mu} = \Lambda^{\mu}_{\ \nu} \, {\tilde y}^{\nu}$ where $\Lambda^{\mu}_{\ \nu}$ is the requisite Lorentz transformation to rotate the direction of motion.  
However, while straightforward, this is still a bit messy, so we will instead proceed using the following trick: we observe that the ``transverse null plane" defined by $y_+ = 0$ in the coordinates of \req{LorAdS} has its spatial cross sections given by a co-dimension 2 hypersurface generated by `zero-energy' spacelike geodesics emanating transverse to the $y_1$ direction.
This is however a purely geometric statement, which therefore applies in any coordinate system.  Since the original form of the metric \req{LorAdS} does not single-out any preferred direction, we can use this as a general prescription for finding any GSW in AdS.

\paragraph{Geodesics in AdS:}
The actual construction of the GSW sourced by a given string bit moving relativistically in direction of $\transv$ therefore proceeds as follows.\footnote{
Here it turns out to be more convenient to use Cartesian (rather than spherical polar) coordinates $(t,x,y,z,u)$, so that the AdS metric is $ds_{\rm AdS}^2 = \frac{1}{u^2} \left[ -dt^2 + dx^2 + dy^2 + dz^2 + du^2 \right]$.}
Without loss of generality, let us fix $t=0$ and consider the support of the GSW from a string bit parameterized by $u$.
A general zero-energy spatial geodesic in AdS has a tangent vector 
$p^a = \dot{x} \, \partial_x^a + \dot{y} \, \partial_y^a + \dot{z} \, \partial_z^a + \dot{u} \, \partial_u^a $, 
and the Killing fields $ \partial_x^a, \partial_y^a$, and $\partial_z^a$ generate constants of motion $K_x \equiv p_a \, \partial_z^a = \frac{\dot{x}}{u^2}$, $K_y= \frac{\dot{y}}{u^2}$ and $K_z= \frac{\dot{z}}{u^2}$.
Defining 
$K^2 \equiv K_x^2 + K_y^2 + K_z^2 $, we can write
the radial part of the geodesic equation as 
\begin{equation}
 p_a \, p^a = 1= \frac{K^2 \, u^4+\dot{u}^2}{u^2} \ ,
\label{}
\end{equation}	
from which we can readily extract the solution for $x(u), y(u), z(u)$ along such a geodesic.
It is easy to verify that all geodesics will have the shape of a semi-circle in the $(x,y,z,u)$ space, intersecting the boundary $u=0$ orthogonally.  The position, orientation, and size of this semi-circle will be given by the initial position and velocity (and the latter is given by the constants of motion $K_i$).
In particular, let $p_i \equiv (x_i, y_i, z_i,u_i)$ denote the initial position, $p_0 \equiv (x_0, y_0, z_0,0)$ denote the endpoint of the (upward part of) the geodesic at $u=0$, and for completeness let $p_m \equiv (x_m, y_m, z_m,u_m)$ denote the symmetric point along the geodesic, which has the largest value of $u$.  We can then write the $x(u)$ part of the geodesic compactly as
\begin{equation}
K^2 \, u^2 = 1 - \frac{K^4}{K_x^2} \, (x-x_m)^2 
 =  K^2 \, \frac{(x-x_0)}{K_x} \, \left( 2 - K^2 \, \frac{(x-x_0)}{K_x}  \right)
\label{geodprof}
\end{equation}	
and similarly for the $y(u)$ and $z(u)$ parts.

Now that we have the general form of a zero-energy geodesic, we can find the specific set of geodesics pertaining to a given string bit by specifying the initial position and velocity.
The initial position $p_i = (x_i, y_i, z_i, u_i)$ is obtained from the string's profile \req{string} in Cartesian coordinates,
$x_i(u_i) = R(u_i) \, \cos[\phi(u_i)] $, 
$y_i(u_i) = R(u_i) \, \cos[\phi(u_i)] $, and
$z_i(u_i) = 0$. 
From each initial position $p_i$ we have a full two-parameter family of geodesics, constrained by the condition of emanating orthogonally to $\transv(u_i)$.
Denoting the initial velocity $\vec{w}_i = (w_x, w_y, w_z, w_u)$, we can  freely specify the components $ w_x$ and $w_y$, and impose the orthogonality condition by 
\begin{equation}
w_u = -R' \, (c \, w_x + s \, w_y) - \frac{R'^2+1}{R \, \phi'} \, (s \, w_x - c \, w_y) 
\label{}
\end{equation}	
and then normalize this velocity by fixing $w_z$ in terms of $ w_x$ and $w_y$ appropriately.
As discussed in \App{a:stress}  below, the dominant contribution to $\CE$ arises from the point along the GSW which is `closest' to the source.  This will occur along the `steepest' geodesic towards the boundary from the point $p_i$, emanating orthogonally to $\transv(u_i)$.  For such geodesic, it is simple to construct $\vec{w}$ by taking any ${\vec {\tilde v}} \propto \transv$ and letting
\begin{equation}
\vec{w}= \sqrt{\frac{{\tilde v}_u^2}{({\tilde v}_x^2+{\tilde v}_y^2) \, ({\tilde v}_x^2+{\tilde v}_y^2+{\tilde v}_u^2)}} \
\left( 0 \ , \ {\tilde v}_x \ , \ {\tilde v}_y \ , \ 0 \ , \
- \frac{{\tilde v}_x^2+{\tilde v}_y^2}{{\tilde v}_u} \right) \ .
\label{wshortest}
\end{equation}	
Finally, from the initial position specified by $u_i$ and a velocity $\vec{w}_i$ we can extract the constants of motion,
$K_x = \frac{w_x}{u_i}$, $K_y = \frac{w_y}{u_i}$, and $K_z = \frac{w_z}{u_i}$,
and relate the endpoint $p_0$ to the initial position $p_i$ via \req{geodprof},
\begin{equation}
x_0 = x_i + \frac{K_x}{K^2} \,  \left( 1 - \sqrt{1 - K^2 \, u_i^2} \right) \ ,
\label{endp}
\end{equation}	
and similarly for $x \to y,z$.

\paragraph{Asymptotic analysis:}
From the asymptotic analysis indicated in \sec{s:asymp}, we can extract these constants of motion explicitly as 
\begin{equation}
K_x = \frac{v^2}{\sqrt{v^4 \, \gamma^2 \, u^2 +R_0^2}} \,
\frac{v^3 \, \gamma \, u \, c  - R_0 \, s}{\sqrt{v^6 \, \gamma^2 \, u^2 +R_0^2}} \ , \qquad
K_y = \frac{v^2}{\sqrt{v^4 \, \gamma^2 \, u^2 +R_0^2}} \,
\frac{v^3 \, \gamma \, u \, s  + R_0 \, c}{\sqrt{v^6 \, \gamma^2 \, u^2 +R_0^2}}\qquad
\label{Kxtas}
\end{equation}	
where 
\begin{equation}
c=\cos \left(\frac{\pi}{2} - \frac{ v \, \gamma \,  u }{R_0} \right)
\ands
s=\sin \left(\frac{\pi}{2} - \frac{ v \, \gamma \,  u }{R_0} \right)
\label{}
\end{equation}	
from which it follows that
\begin{equation}
K^2 = \frac{v^4}{v^4 \, \gamma^2 \, u^2 +R_0^2} \ .
\label{Ksqas}
\end{equation}	

\paragraph{Regularized proper length}
Finally, let us find the (regularized) proper length along such a geodesic.  
The direction of the geodesic is unimportant, only the starting point $p_i$ and the total momentum $K$ matters; these determine how far from $p_i$ is the final point $p_0$.
Since the proper length along such a geodesic, between $u_i$ and some final value $u_f$ is 
\begin{equation}
L= \ln \left( \frac{u}{1+\sqrt{1 - K^2 \, u^2}} \right) \arrowvert_{u_i}^{u_f}
\label{}
\end{equation}	
we see that the $u_f \to 0$ limit generates a universal piece $-\ln u/2 \to \infty$ which can be regulated away, by the now standard procedure.  Rewriting $K$ in terms of the geodesic endpoint on the boundary we obtain the regularized proper length
\begin{equation}
\CL(d;u_i) = \ln \left( \frac{u_i^2 + d^2}{2 \, u_i} \right) 
\qquad {\rm where}\ \ 
d^2 \equiv (x_f - x_i)^2 + (y_f - y_i)^2 + (z_f - z_i)^2 \ .
\label{}
\end{equation}	
%

\section{Boundary stress tensor induced by GSW}
\label{a:stress}

In this Appendix we wish to determine the actual profile of the boundary stress tensor induced by a given GSW. 
We make use of the calculation performed already in \cite{Gubser:2008pc}.\footnote{
In fact, these authors consider a more complicated situation involving collisions of gravitational shock waves (see also \cite{Grumiller:2008va} and earlier work of \cite{Nastase:2004pc}) in context studying the entropy production in such a collision in order to compare with the entropy production in heavy ion collisions.  Since before collision the waves superpose linearly, here we will simply ignore one of the waves. 
We also rewrite their results in our notation.}
A GSW sourced by a point particle propagating {\it parallel} to the boundary in the $x$ direction at $u=1$, with $y=z=0$ and $x=y$ has the following metric:
\begin{equation}
ds^2 = \frac{1}{u^2} \, \left[ -dt^2 + dx^2 + dy^2 + dz^2 + du^2 \right] 
+\frac{1}{u} \, \Phi(y,z, u) \, \delta(t-x) \, (dt-dx)^2 
\label{parGSWmet}
\end{equation}	
where, up to an overall normalization factor,
\begin{equation}
\Phi(y,z,u) \propto \frac{1+8q \, (1+q)-4\, (1+2q) \, \sqrt{q\, (1+q)}}{\sqrt{q\, (1+q)}} \ , 
\qquad
q \equiv \frac{y^2 + z^2 + (u-1)^2 }{4 \, u} \ .
\label{Phiq}
\end{equation}	
Note that the source of the GSW is at $q=0$.
We can easily generalize this set-up to a particle propagating parallel to the boundary at any other bulk radial distance $u= u_i$ by using symmetry of AdS. Rescaling coordinates 
$t \to \alpha \, t$, $x \to \alpha \, x$, $y \to \alpha \, y$, $z \to \alpha \, z$, and $u \to \alpha \, u$, 
leaves the form of the metric \req{parGSWmet} invariant, and 
$\Phi(\alpha \, y, \alpha \, y, \alpha \, y)$ likewise has the same functional dependence on $q$ in \req{Phiq}, but this function is now different:
\begin{equation}
q = \alpha\, \frac{y^2 + z^2 + (u-\frac{1}{\alpha})^2 }{4 \, u}
\label{qrescaled}
\end{equation}	
In particular, the source of the GSW at $q=0$ propagates at $u_i = 1/\alpha$.

Now that we have the exact metric of a GSW for a particle propagating parallel to the boundary at any given $u = u_i$ in Fefferman-Graham coordinates, it is straightforward to read off the induced stress tensor on the boundary in the standard manner \cite{deHaro:2000xn}.   
Since for a metric of the form
\begin{equation}
ds^2 = ds_{AdS}^2 + \frac{1}{u^2} \, \delta g_{\mu \nu} \, dx^{\mu} \, dx^{\nu} \ ,
\label{genmetdef}
\end{equation}	
the boundary stress tensor expectation value 
is given by
\begin{equation}
\vev{T_{\mu \nu}} \propto \lim_{u\to 0} \frac{1}{u^4} \, \delta g_{\mu \nu} \ ,
\label{Tfromgenmetdef}
\end{equation}	
we find that in the present case \req{qrescaled} the boundary energy density ${\cal E} $ sourced by a null particle at $u=u_i$ is given by 
\begin{equation}
{\cal E}(t,x,y,z) = \vev{T_{tt}} 
\propto \frac{\delta(t-x)}{\alpha^3 \, \left( \frac{1}{\alpha^2} + y^2 + z^2 \right)^3 } 
=\frac{\delta(t-x) \, u_i^3}{\left(u_i^2 + y^2 + z^2 \right)^3 }  \ .
\label{EGSWplane}
\end{equation}	

Using the results in \App{a:GSWconstr}, we can recast this expression in terms of a regularized proper length along a geodesic from the source towards the boundary along the GSW (in this case along constant $x$).  In particular, writing this more generally in terms of a geodesic between the source at $p_i$ and a boundary point $p_0$, with regularized proper length $\CL(p_i, p_0)$, the profile of energy density (which is supported along the boundary value of the GSW) is 
\begin{equation}
{\cal E}(p_0) \propto  
 e^{-3 \, \CL(p_i, p_0)} \ .
\label{EGSWplaneL}
\end{equation}	
In \sec{s:results} we will take this expression to be valid in full generality, even when the source is not propagating parallel to the boundary.  Note, however, that in the interesting case of a relativistic quark for which $v \, \gamma \gg1$, the source indeed does propagate approximately parallel to the boundary.
}

\providecommand{\href}[2]{#2}\begingroup\raggedright\endgroup

\end{document}